\documentclass[a4paper,11pt]{article}
\pdfoutput=1 
\usepackage{jcappub} 
\usepackage{amsmath}
\usepackage{amssymb}
\usepackage{float}
\usepackage{appendix}
\usepackage[T1]{fontenc} 

\def\ds{{\rm d}s}
\def\da{{\rm d}a}

\title{\boldmath Statistical Isotropy of the CMB \texorpdfstring{$E$}{E}-mode signal}

\author[a,1]{Joby P. Kochappan \note{Corresponding author}}
\author[b]{, Aparajita Sen}
\author[a]{, Tuhin Ghosh}
\author[c]{, Pravabati Chingangbam}
\author[b]{, Soumen Basak}

\affiliation[a]{School of Physical Sciences, National Institute of Science Education and Research, HBNI, Jatni 752 050, Odisha, India}
\affiliation[b]{School of Physics, Indian Institute of Science Education and Research Thiruvananthapuram, Maruthamala PO, Vithura, Thiruvananthapuram 695551, Kerala, India}
\affiliation[c]{Indian Institute of Astrophysics, Koramangala 2 Block, Bangalore 560 034, India}

\emailAdd{joby@niser.ac.in}
\emailAdd{aparajita15@iisertvm.ac.in}
\emailAdd{tghosh@niser.ac.in}
\emailAdd{prava@iiap.res.in}
\emailAdd{sbasak@iisertvm.ac.in}

\abstract{We test the statistical isotropy (SI) of the $E$-mode polarization of the cosmic microwave background (CMB) radiation observed by the Planck satellite using two statistics, namely, the contour Minkowski Tensor (CMT) and the Directional statistic ($\mathcal{D}$ statistic). The parameter $\alpha$ obtained from the CMT provides information of the alignment of structures and can be used to infer statistical properties such as Gaussianity and SI of random fields. The $\mathcal{D}$ statistic is based on detecting preferred directionality shown by vectors defined by the field. These two tests are complementary to each other in terms of sensitivity at different angular scales. The CMT is sensitive towards small-scale information present in the CMB map while $\mathcal{D}$ statistic is more sensitive at large-scales. We compute $\alpha$ and $\mathcal{D}$ statistic  for the observed $E$-mode of CMB polarization, focusing on the SMICA maps, and compare with the values calculated using FFP10 SMICA simulations  which contain both CMB and noise. We find good agreement between the observed data and simulations. Further, in order to specifically analyze the CMB signal in the data, we compare the values of the two statistics obtained from the observed Planck data with the values obtained from isotropic simulations having the same power spectrum, and from SMICA noise simulations. We find no statistically significant deviation from SI using the $\alpha$ parameter. From $\mathcal{D}$ statistic we find that the data shows slight deviation from SI at large angular scales.} 

\begin{document}
\maketitle
\flushbottom
\section{Introduction}
\label{sec:intro}

 The $\Lambda$CDM model has been very successful at explaining the cosmological observations to date, and is currently the most widely accepted model of the Universe \cite{wmap9,planck18:i}. Two important 
2
 assumptions of this model are that the Universe is homogeneous and isotropic at large-scales. The validity of these crucial assumptions have been subjected to various tests using cosmological data such as the cosmic microwave background (CMB) radiation and the large-scale structure of the distribution of matter in the universe. Our goal is to test the statistical isotropy (SI) of the Universe using observations of CMB polarization made by the Planck satellite. The CMB polarization field can be split into two components, the curl-free component called the $E$-mode, and the divergence-free component called the $B$-mode. For this work, we focus on the SI of the $E$-mode of CMB polarization observations, following up our recent work \cite{Joby:2019} where we had analyzed the SI of CMB temperature anisotropies.  
 
It is important to test SI using different approaches that are complementary to each other for obtaining robust constraints. In the literature, various methods can be found for testing the SI of CMB data. Hajian et al. \cite{Hajian:2004,Basak:2006,Tuhin:2007} formulated the BiPolar Spherical Harmonics (BiPoSH) technique to test the isotropy of CMB maps. They applied the BiPoSH technique to the WMAP 3-year data, and found no significant violation of SI in the temperature maps, but $\simeq 2\sigma$ deviation from SI \cite{Souradeep:2006} in the $E$-mode map. Another technique based in harmonic space is the power tensor method \cite{Pranati:2017}, where the eigenvalues of the power tensor constructed from the coefficients of the spherical harmonics contain information about the SI of the field. Using this method, the authors do not find any significant deviations from SI in the temperature maps of the WMAP 3-year data and Planck 2013 data. More recently, the method of multipole vectors was applied to the Planck 2015 and Planck 2018 temperature data, and the data was found to be in good agreement with the assumption of SI \cite{Renan:2019}. Eriksen et al. \cite{Eriksen:2007} performed a Bayesian analysis on WMAP 3-year data to find $ > 2\sigma$ evidence for the presence of a hemispherical power asymmetry in the temperature maps. The local variance estimator method \cite{Shafieloo:2017}, is based on real space fields and was applied to the $E$-mode Commander map of the Planck 2015 data. The authors found $3\sigma$ evidence for hemispherical power asymmetry at large angular scales.
 
 
 Minkowski Tensors (MTs) \cite{McMullen:1997,Alesker:1999,Hug:2008,Schroder3D:2013,mtsphere:2017} carry information on the shapes of structures. In the context of the CMB, the word structure here refers to hotspots and cold spots defined by iso-field contours. One of the MTs, the contour MT (CMT) has been used to test the SI of random fields ~\cite{Joby:2019,Vidhya:2016,mtsphere:2017,stephen:2018,stephen:2019}. This method was first applied in~\cite{Vidhya:2016} to the Planck 2015 data release. The authors found no significant violation of SI in the CMB temperature field, but obtained $\simeq$ $4\sigma$ deviation for CMB $E$-mode field. The calculation of the contour MT relied on first carrying out stereographic projection of the field on the two dimensional plane, and then constructing the iso-field contours. Stereographic projection can introduce numerical error into the estimation of the isotropy parameter, and in~\cite{mtsphere:2017} a new method for the calculation based on field derivatives directly on the sphere was developed. This new method of estimating MTs was used in \cite{Joby:2019} to test SI in the temperature data of the Planck 2018 data release. The polarization part of the  Planck 2018 data has been significantly improved in terms of removal of systematics, foreground modelling and instrumental noise reduction as compared to the Planck 2015 data release\cite{planck18:i,planck18:ii,planck18:iii}. This paper addresses the test of SI of the Planck 2018 polarization data \footnote{http://pla.esac.esa.int/pla}, in particular $E$-mode using the method of calculation on the sphere.  

 To complement our results from CMTs we use a second test for SI, the Directionality test or the $\mathcal{D}$ statistic \cite{Bunn and Scott}. It is a statistical test which has been devised to measure any preferred directionality over the sky. The test follows a simple algorithm making it numerically inexpensive. It was first applied in  ~\cite{Bunn and Scott} to the COBE-DMR temperature data. Subsequently, the test has also been applied to WMAP and Planck 2018 polarization data in ~\cite{D.Hanson et al} and ~\cite{D.Ghrear et al}. In ~\cite{D.Ghrear et al} the authors have constructed a pseudo vector constructed from the $Q$ and  $U$ maps to demonstrate the absence of residual foreground in  the cleaned (and masked) CMB maps. 
 
 This article is organized as follows. All the data pre-processing steps that we follow before applying the SI tests have been described in section ~\ref{sec:data}. Section ~\ref{sec:methods} presents an outline of the MFs and $\mathcal{D}$ statistic methods.  It also carries a description of how the CMTs can be estimated from pixelated maps of random fields on the sphere. In section ~\ref{sec:noisy}, we discuss how the addition of an isotropic white noise component to a given anisotropic signal map affects the $\alpha$ and $\mathcal{D}$ values of the resultant map. Section \ref{sec:results} presents the main results of this paper, comparisons of the $\alpha$ and $\mathcal{D}$ statistic  estimated from the observed SMICA $E$-mode map with those from the FFP10 SMICA simulations, isotropic simulations and SMICA noise simulations. Finally, in section ~\ref{sec:conc}, we draw conclusions based on our results. We have also demonstrated that $\mathcal{D}$ statistic is more sensitive at larges scales compared to CMTs in Appendix ~\ref{App:HA}.


\section{Data}
\label{sec:data} 
2

In this section, we describe the set of observed data and simulations that we have used for our analysis. We also define the notations used to denote the various sets of maps that we work with.

\subsection{Planck data and mask}
We use the publicly available Planck 2018 data release SMICA $I$,$Q$,$U$ maps from the Planck Legacy Archive (PLA). These maps are provided at a beam resolution of $5'$ full-width at half maximum (FWHM) and projected on HEALPix\footnote{http://healpix.sourceforge.net} pixel resolution of $N_{\rm side}=2048$ \cite{planck18_iv,planck18_vii}. These maps combine multi-frequency sky observations by the Low Frequency Instrument (LFI, $30-70$\,GHz) and the High Frequency Instrument (HFI, $100-857$\,GHz). The HFI has an angular resolution of $\approx 5'-10'$ FWHM, while the LFI has a resolution of $\approx 13'-33'$ FWHM. For the polarization data, scalar Minowski functionals (and also MTs) computed from the spin-2 $Q$ and $U$ Stokes parameters are not invariant under rotations after masking \cite{Chingangbam:2017}. Therefore, we convert the full sky SMICA $Q$ and $U$ maps to $E$- and $B$-mode maps and use only the $E$-mode map for our analysis. 

\begin{figure}
\centering
\resizebox{5.2in}{4.2in}{\includegraphics{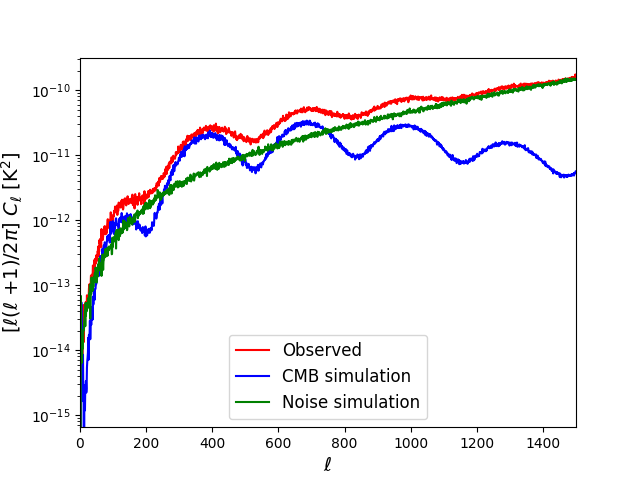}}
\caption{Comparison of $EE$ power spectra estimated from the masked observed SMICA map (red line), SMICA CMB simulation (blue line) and SMICA noise simulation (green line). The excess power seen in the observed SMICA map is due to the instrument noise. All maps have pixel resolution given by $N_{\rm side}=2048$ and effective beam size of $5'$ FWHM.}
\label{fig:power_spectra}
\end{figure}

For the analysis of Planck polarization data, the recommended mask is the common polarization mask, which is available at a resolution of $N_{\rm side}=2048$ and has a sky fraction of $\approx 77\%$. We apply the MASTER code \cite{master} to compute $EE$ power spectrum of SMICA map after applying the common polarization mask. We find that the noise present in the SMICA $E$-mode map dominates over the expected CMB $E$ mode signal from the Planck best-fit $\Lambda$CDM model at multipoles $\ell > 700$ as will be described in section \ref{sec:noise_level} (see figure ~\ref{fig:power_spectra}). We first smoothed the SMICA $E$-mode map with a Gaussian beam resolution of $20'$ FWHM and downgraded it to a HEALPix pixel resolution of $N_{\rm side}=512$. The map pixel resolution is chosen such a way that the effective beam smoothing is done over two to three pixels. This is necessary to get reliable estimates of the derivatives of the field. We will refer to the smoothed SMICA $E$-mode map as the {\it observed $E$ map}. We also smooth and downgrade the common polarization mask to $20'$ FWHM and $N_{\rm side}=512$. We apply this smoothed mask to the observed $E$ map before computing the statistics for testing SI.

\subsection{Simulations}

We use the Full Focal Plane (FFP10) SMICA $I$,$Q$,$U$ simulations, which include CMB and noise components, to make a comparison with the results obtained from the observed $E$ map. These simulated maps have the same beam resolution as the observed SMICA $IQU$ maps. The SMICA CMB simulations contain only the CMB component convolved with $5'$ FWHM Gaussian beam, while the SMICA noise simulations contain only the instrumental noise contribution. Similar to the observed SMICA map, we convert the simulated $Q$ and $U$ maps to $E$ and $B$ maps and then work with the $E$-mode maps. These simulated SMICA CMB and noise maps are also smoothed with a Gaussian beam of $20'$ FWHM, downgraded to pixel resolution $N_{\rm side}=512$, and masked with the common polarization mask. 

To study the effect of isotropic noise on $\alpha$ and $\mathcal{D}$ statistic, we use the foreground model $I$,$Q$,$U$ simulations at 353 GHz that are publicly available at the PLA. We convert them into the respective $E$- and $B$-mode maps and then use the $E$-mode maps for our analysis. These foreground maps are given at a resolution of $N_{\rm side}=2048$ and a beam of $5'$ FWHM. Again, we downgrade the pixel resolution of these maps to $N_{\rm side}=512$, smooth them with a Gaussian beam of $20'$ FWHM and mask them with the common polarization mask. 

To infer the SI properties of the CMB $E$-mode signal, we make a set of 300 isotropic simulations. First we make a set of isotropic CMB $E$-mode simulations using the best-fit $\Lambda$CDM $EE$ power spectrum \cite{planck18_v} convolved with the pixel window function for $N_{\rm side}=512$ and a Gaussian beam of $20'$ FWHM.  Using the mean noise $EE$ power spectrum from 300 SMICA noise simulations, convolved with the pixel window function of $N_{\rm side}=512$ and beam of $20'$. Then we add the two sets of simulations to obtain 300 isotropic simulations having the same power spectrum as the observed $E$ map.

\subsection{Level of noise in the observed \texorpdfstring{$E$}{E} map}
\label{sec:noise_level}

It is well known that the signal-to-noise ratio of the observed SMICA $E$ map is of order one, and is lower than the corresponding SMICA temperature map \cite{planck18_iv}. The SI properties of a low signal-to-noise observed $E$ mode map can be significantly affected by the noise properties. For example, if the signal is anisotropic by nature and is dominated by isotropic noise, then the resultant map can become statistically isotropic, and vice versa. For this reason, it is important to check the noise level in the observed $E$ map before drawing conclusions on the SI of the CMB $E$-mode signal.

In figure ~\ref{fig:power_spectra}, we compare the scaled power spectrum, $D_{\ell}=\ell(\ell+1)C_{\ell}/2\pi$, where $C_{\ell}$ is the angular power spectrum, of the observed SMICA $E$ map with that of a corresponding SMICA simulation and SMICA noise simulation. The red line represents the observed SMICA $E$ map, which contains the $E$-mode signal, residual foreground and the noise. The SMICA CMB simulation contains the CMB $E$-mode signal and is represented by the blue line, while the SMICA noise simulation is represented by the green line. From figure ~\ref{fig:power_spectra}, we note that the SMICA $E$ map is noisy across all scales, and hence it is important to understand the effect of the instrument noise on the SI of the observed $E$ map.


\section{Quantifying the Statistical Isotropy of CMB maps}\label{sec:methods}
In this section, we describe the two statistical methods that we use to test for SI of the observed $E$ map, namely, MTs and $\mathcal{D}$ statistic. We start with the definitions of the two statistics and proceed to discuss how they capture the SI information of a given random field.

\subsection{Minkowski Tensors}
\label{sec:mt_on_sphere}

First, we give a brief overview of the Contour Minkowski Tensor (CMT) which is our primary tool for testing SI. Let $C$ be a closed curve on the unit sphere.   The CMT associated with this curve, denoted by $\mathcal{W}_1$, is defined as,
\begin{equation}
    \mathcal{W}_1 = \frac{1}{4}\int_C \hat{T} \otimes \hat{T} \, \ds ,
    \label{eqn:cmt}
\end{equation}
where the integral is over $C$, $\hat{T}$ denotes the unit tangent vector to the curve at each point, $\otimes$ denotes the symmetric tensor product of two vectors, and $\ds$ denotes the infinitesimal arc length line element along the curve. For a smooth random field, denoted by $u$, excursion sets consist of points on the sphere where the field has values higher than some chosen threshold value, which we denote by $\nu_t$. The boundaries of the excursion sets, indexed by $\nu_t$, form closed curves. For multiple curves, $\mathcal{W}_1$ can be obtained as the sum over all the curves per unit area of the unit sphere (note that we use the same notation as for the single curve). In order to compute it we can convert the line integral to an area integral using a suitable Jacobian \cite{Schmalzing:1998}, and $\mathcal{W}_1$ can be expressed in terms of the field and its derivatives as,
\begin{equation}
    \mathcal{W}_1 = \frac{1}{16\pi}\int_{S^2} \da \, \delta \left( u-\nu_t \right) \, \left| \nabla u \right| \, \hat{T} \otimes \hat{T},
    \label{eqn:cmt_area}
\end{equation}
where, $\da$ is the area element on the sphere, $\delta$ is the Dirac delta function, and $\nabla$ denotes covariant derivative on the sphere. The suffix $S^2$ indicates that the integral is over the unit sphere. $\mathcal{W}_1$ can be re-expressed in terms of the field, $u$, and it's derivatives, $u_{;i}$, as,
\begin{equation}
\mathcal{W}_{1}=\frac{1}{16\pi}\int_{S^2} \da \, \delta(u-\nu_{t}) \, \frac{1}{\left|\nabla u\right|} \, \mathcal{M},
\label{eqn:w1_num}
\end{equation}
where $\mathcal{M}$ is a symmetric matrix and is given by
\begin{equation}
\mathcal{M}=\begin{pmatrix}u_{;2}^{2} & u_{;1}u_{;2}\\
u_{;1}u_{;2} & u_{;1}^{2}
\end{pmatrix}. \label{eq:mat_m}
\end{equation}

Choosing to label the eigenvalues of $\mathcal{W}_1$ as $\Lambda_1$ and $\Lambda_2$ such that $\Lambda_1 \leq \Lambda_2$, the alignment parameter, $\alpha$, is defined as the ratio of the two eigenvalues \cite{mtsphere:2017},
\begin{equation}
    \alpha = \frac{\Lambda_1}{\Lambda_2} \ .
\end{equation}
$\alpha$ gives a measure of the alignment of the structures in the level sets, which reflects the SI of the random field. If the field is statistically isotropic, then the value of $\alpha$ will be close to unity. If the field is not statistically isotropic, then the value of $\alpha$ will deviate from unity and move towards zero. For an isotropic Gaussian random field, the ensemble expectation value of $\mathcal{W}_1$ has been calculated analytically \cite{mtsphere:2017} and is given by,

\begin{equation}
\centering
   \left< \mathcal{W}_1 \right>_{ij} = \left\{ \begin{tabular}{cl}
     $\frac{1}{16\sqrt{2}r_c}\exp\left( -\nu^2/2 \right)$ & \text{if $i$=$j$}, \\
     & \\
     0 & \text{otherwise}.
   \end{tabular}
   \right.
   \label{eqn:cmt_expectation}
\end{equation}
In the above expression,  $\nu\equiv\nu_t/\sigma_0$ is the field threshold in units of the standard deviation of $u$, denoted by $\sigma_0$. The quantity $r_c$ represents a characteristic scale of the size  of structures in the field and is defined by $r_c\equiv\sigma_0/\sigma_1$, where $\sigma_1$ is the standard deviation of the gradient of the field.   
In eq. \ref{eqn:cmt_expectation} the eigenvalues are equal and the alignment parameter is $\alpha$ = 1. Since we have chosen to label the eigenvalues such that $\Lambda_1 \, \leq  \Lambda_2$, we have $0 \, \leq \alpha \leq 1$. If there is a preferred direction in the sky that causes an alignment of structures in the level sets, the value of $\alpha$ will be much lower than 1. 

In the numerical implementation of eq.~\eqref{eqn:cmt_area} to compute the CMT and $\alpha$ from a discretely sampled random field on the sphere, the Dirac $\delta$ function is replaced by the following approximation \cite{Schmalzing:1998},

\begin{equation}
\centering
    \delta \left( u - \nu_t \right) = \left\{ \begin{tabular}{cl}
         $1/(\Delta \nu_t)$ & \text{ if } $\nu_t - \Delta \nu_t /2 < u < \nu_t + \Delta \nu_t/2$, \\
         & \\
         0 & \text{ otherwise }.
    \end{tabular} \right.
    \label{eqn:delta}
\end{equation}

As discussed in \cite{mtsphere:2017}, using the expression for $\left< \mathcal{W}_1 \right>$ given in eq.~\eqref{eqn:cmt_area} and the approximation given in eq.~\eqref{eqn:delta}, we can numerically compute the CMT and $\alpha$ for a pixelated map on the sphere. We will work with the field rescaled by its standard deviation, and hence our results for $\alpha$ will be presented as functions of $\nu$.
The approximation in eq.~\eqref{eqn:delta} has inherent numerical inaccuracy~\cite{Lim:2012}. However, the numerical errors for the two eigenvalues are comparable, and hence get cancelled out when we compute $\alpha$, as shown in~\cite{Goyal:2019vkq}. Therefore, this method is very well suited for using $\alpha$ to test SI.

\subsection{Directionality Test}
\label{sec:D-Stat Theory}

In this section we will briefly describe the basic concepts behind the Directionality test for the CMB which was first introduced in ~\cite{Bunn and Scott}. It is a statistical test defined in pixel domain to check for any preferred directionality present in the observed CMB signal. To achieve this goal a vector at each point in the sky is defined to capture the directional properties of the field. For a scalar field the vector can be simply the gradient of $u$. The alignment of these gradients towards any particular direction is quantified by its projection in the direction of the sky. Mathematically, this can be implemented by defining a function $f$ for each direction $\hat{\boldsymbol{n}}$ in the sky,
\begin{equation}
    f(\hat{\boldsymbol{n}})=\sum_{p=1}^{N_{\rm pix}}w_{p}\left(\hat{\boldsymbol{n}}.\vec{\nabla}u\right)^{2}\label{eq: f_n_cap},
\end{equation}
where $p$ stands for the pixel index of the map, $w_p$ are  weights for each pixels on the map, $N_{\rm pix}$ is the total number of pixels of the map .
If there is no preferred directionality present in the sky, all the values of $f(\hat{\boldsymbol{n}})$ should be close to each other. Any anomalously large or small values of $f(\hat{\boldsymbol{n}})$  will indicate the presence of directionality in the data. This information is captured by the $\mathcal{D}$ statistic which is defined as,
\begin{equation}
    \mathcal{D}=\frac{{\rm minimum} [f(\hat{\boldsymbol{n}})]}{{\rm maximum} [f(\hat{\boldsymbol{n}})]}.\label{eq:D-ratio}   
\end{equation}

For an isotropic field the values of $\hat{\boldsymbol{n}}$ for different directions are expected to be close to each other because no particular direction is favoured. This leads to the value of $\mathcal{D}$ being close to unity. So, any anisotropic feature present in the map will manifest itself as deviations of $\mathcal{D}$ from unity.Note that $\mathcal{D}$ defined by eq. \eqref{eq:D-ratio} is the inverse of the  $\mathcal{D}$-ratio in ~\cite{Bunn and Scott} . It takes values in the range $[0,1]$.

The $\mathcal{D}$ statistic is applicable to partial sky as well. However, the masked region of the sky will make an anomalous directional contribution to $\mathcal{D}$ statistic. So, in order to mitigate the impact of partial sky on the measurement of $\mathcal{D}$ statistic, non-uniform weights are assigned to the pixels in the unmasked region of the map. 
The weights are determined based on two assumptions. The first assumption is that the field is isotropic. For such cases, the ensemble average of $f(\hat{\boldsymbol{n}})$ should be directionally independent. However, this gives rise to only six independent equations which are insufficient to determine all the pixel weights. The second assumption is that the for an isotropic case the variation of signal from pixel to pixel should not be drastically different. Hence, to determine the pixel weights the variance of weights is also minimized. Based on these assumptions, the functional form of $w_{p}$ is described by the Cartesian components of the pixel centres, $\hat{\boldsymbol{r}}_{p}$ and the gradient vectors of the scalar field, $P_p$.
\begin{equation}
    w_{p}=w_{p}\left(P_{p}^{-1},r_{pij}\right)\label{eq: D_weights},
\end{equation}
where,
\begin{equation}
    P_{p}=\frac{1}{2}\left\langle \vec{\nabla}u.\vec{\nabla}u\right\rangle. \label{eq:D_P_p}
\end{equation}

It is important to note that the pixel weights are derived from the ensemble average of gradient vectors. Thus, to get an accurate estimate of the weights many realizations of the sky consistent with the data are required. As mentioned in \cite{Bunn and Scott}, for the simulations of the sky, which is  isotropic in nature, the variation of the values of $P_p$ across the sky is very small. However, the situation is different when the sky is not isotropic. If the simulations are made by co-adding inhomogeneous noise with an isotropic signal, the values of $P_p$ will show some variation across the sky. In such situations, noisier patches of the sky will have higher values of $P_p$ and will be down-weighted compared to less noisy patches of the sky. As a consequence, the results obtained using $\mathcal{D}$ statistic will be more sensitive to those patches of sky that are less noisy compared to others. 

In our work, we have computed the pixel weights such that they are independent of any sky simulations. We have done this by assuming all our simulations are isotropic in nature and have set $P_p=1$ for all pixels. Therefore, the weights will only depend on the sky fraction and the morphology of the masked region.  

\section{Effect of adding isotropic noise component to an anisotropic signal}
\label{sec:noisy}

\begin{figure}
    \centering
    \resizebox{3.0in}{2.0in}{\includegraphics{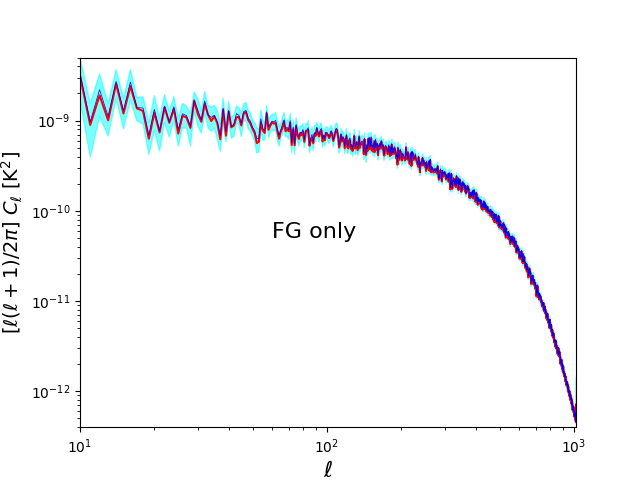}}
    \hskip 0in \resizebox{3.0in}{2.0in}{\includegraphics{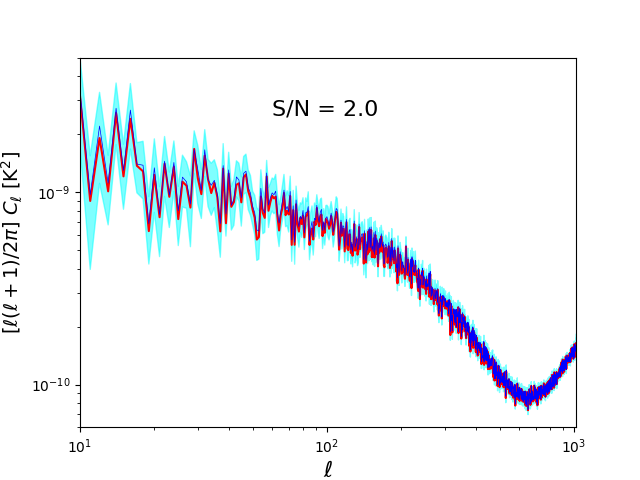}}
    \vskip 0in
    \resizebox{3.0in}{2.0in}{\includegraphics{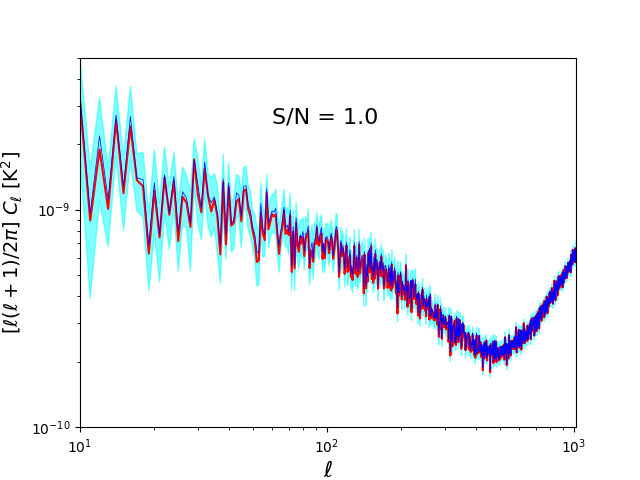}}
    \hskip 0in \resizebox{3.0in}{2.0in}{\includegraphics{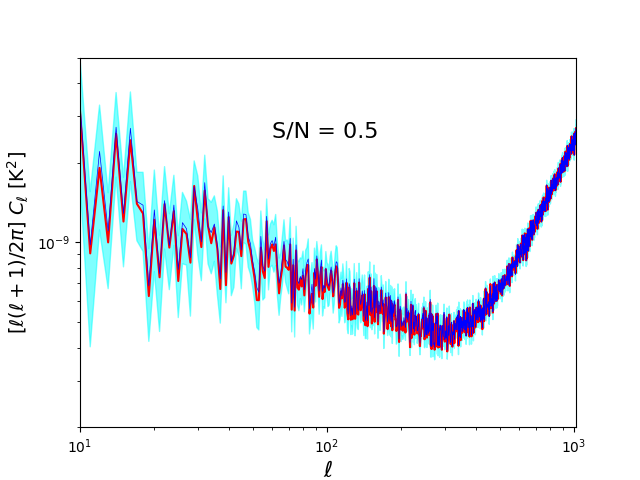}}
    \caption{Comparison of $EE$ power spectra for noisy foreground map (red lines) and mean (blue lines) and $1\sigma$ error bars (cyan shaded regions) obtained from 100 isotropic simulations. The maps having only the foreground component and those with added white noise having S/N ratios of 2, 1 and 0.5, are shown in the \textit{top left}, \textit{top right}, \textit{bottom left} and \textit{bottom right panels} respectively.}
    \label{fig:power_fg+wnoise}
\end{figure}

 To understand the impact of isotropic noise on the SI of the low signal-to-noise map, we make use of the anisotropic foreground model $E$-mode map at 353 GHz. We use the sum of FFP10 thermal dust and synchrotron templates at 353 GHz as our signal (indicated by super script `FG'), and add varying levels of isotropic white noise (indicated by super script `wnoise'). The level of white noise added is given by the inverse of the signal-to-noise ratio (S/N), which we define as the average of $\sqrt{C_{\ell}^{\rm FG}/C_{\ell}^{\rm wnoise}}$, in the multipole range $\ell = 400 - 450$. Here, $C_{\ell}^{\rm FG}$ and $C_{\ell}^{\rm wnoise}$ refer to the $EE$ power spectra of the foreground and white noise simulations, respectively. For our analysis, we choose to add different levels of noise contamination corresponding to S/N = 0.5, 1 and 2, in addition to the case where no noise is added. 
 We then mask these maps with the common polarization mask, compute the pseudo $EE$ power spectra from the masked maps and then use the MASTER code \cite{master} to estimate the full-sky $EE$ power spectra. With these power spectra as the inputs in HEALPix, we generate isotropic foreground plus noise simulations for each of the selected levels of noise contamination. Figure ~\ref{fig:power_fg+wnoise} displays the power spectra of the noisy foreground maps (\textit{red solid line}) and the mean and $1\sigma$ limits from 100 corresponding isotropic simulations (\textit{blue solid line}) for four different S/N ratios. 
 As expected, we can see that with increasing levels of added white noise, the power spectra become dominated by noise at small-scales. 

 We compare the results for $\alpha$ and $\mathcal{D}$ statistic using the noisy foreground maps with the corresponding isotropic simulations, which have the same power $EE$ spectrum. The goal of this exercise is to determine the level of white noise which effectively makes the noisy foreground map isotropic, despite the original foreground signal being anisotropic. 
\subsection{CMT Analysis}

\begin{figure}
    \centering
    \resizebox{3.0in}{2.0in}{\includegraphics{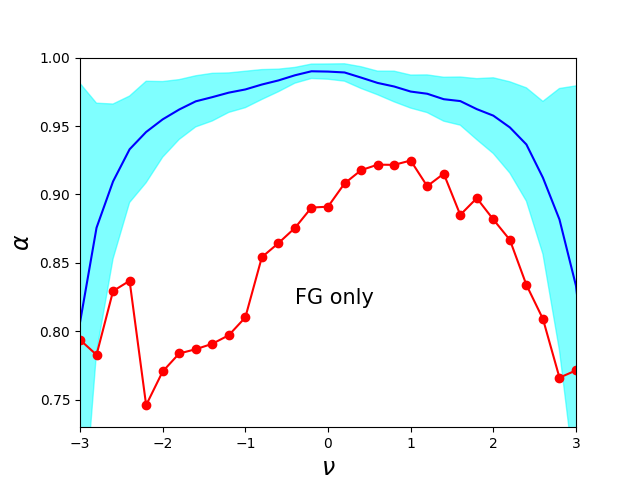}}
    \hskip 0in \resizebox{3.0in}{2.0in}{\includegraphics{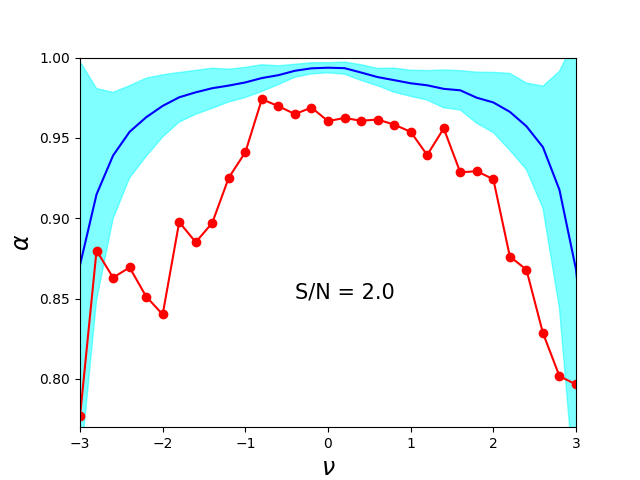}}
    \vskip 0in
    \resizebox{3.0in}{2.0in}{\includegraphics{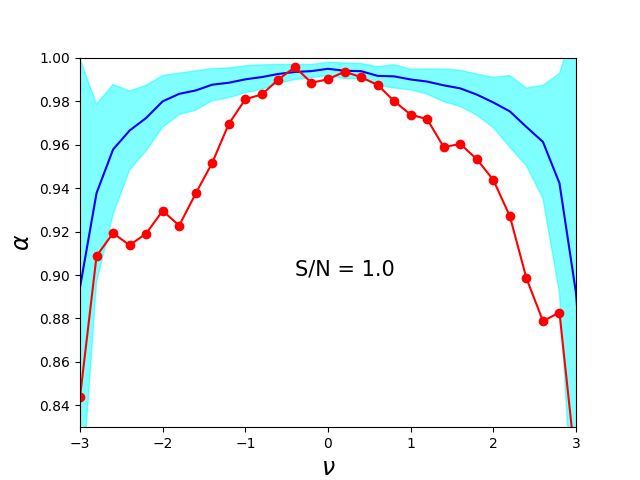}}
    \hskip 0in \resizebox{3.0in}{2.0in}{\includegraphics{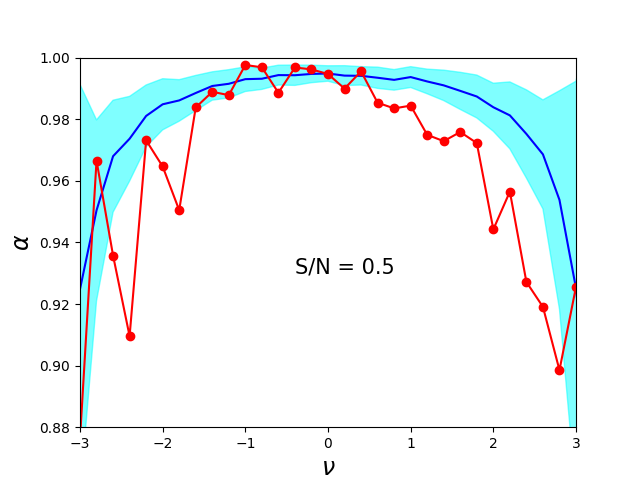}}
    \caption{Comparison of $\alpha$ from noisy foreground maps (red lines) and mean (blue lines) and 1$\sigma$ error bars (cyan shaded regions) from 100 isotropic simulations having the same $EE$ power spectra, for the FG only map (\textit{top left}), and added noise with S/N ratios of 2 (\textit{top right}), 1 (\textit{bottom left}) and 0.5 \textit{bottom right panels}, respectively.}
    \label{fig:353fg+wnoise}
\end{figure}

We present $\alpha$ for the noisy foreground maps with three different S/N ratios of 2, 1 and 0.5 along with the no noise case, in figure ~\ref{fig:353fg+wnoise}. In each panel, the red line represents $\alpha$ values for the noisy foreground map, while the blue line represents the mean from 100 isotropic simulations, and the cyan shaded region represents the 1-$\sigma$ error bars. The top left panel compares the foreground simulations with no white noise, with the isotropic simulations having the same $EE$ power spectrum. There is a clear disagreement between the $\alpha$ values computed from the foreground simulations and those from the isotropic simulations. The top right panel shows the comparison of noisy foreground map having S/N\,=\,2 with the corresponding isotropic simulations. While there is an increase in the $\alpha$ values, there is still a significant difference, $\approx 3\sigma$, between the $\alpha$ values computed from the two sets of maps. For the case with S/N = 1 and 0.5, as shown in the bottom left and right panels, respectively, the difference between the noisy foreground maps and the isotropic simulations is less than $2\sigma$. This means that, for the 353 GHz foreground model $E$-mode map, if isotropic white noise is added at a level greater than S/N = 0.5, $\alpha$ is no longer able to capture the SI properties of the signal. The addition of isotropic noise increases the $\alpha$ values of the total map and makes it effectively isotropic, even though the signal is anisotropic. In the following section, we will use this result to draw conclusions on the SI of the CMB $E$-mode signal from noisy observed $E$-mode map. 

\subsection{\texorpdfstring{$\mathcal{D}$}{D} statistic}
\label{ssec:D-Stat Sensitivity}

Here, we present the results from Directional statistic. In order to quantify the deviations shown by noisy foreground maps from the isotropic simulations we define $\Delta \mathcal{D}$ as following, 
\begin{equation}
\Delta \mathcal{D}=\frac{\left|\mathcal{D}-\bar{\mathcal{D}}_{\rm sims}\right|}{\sigma_{\rm sims}}.
\end{equation}
Where, $\mathcal{D}$ is obtained from the noisy foreground maps, $\bar{\mathcal{D}}_{\rm sims}$ is the mean of and $\sigma_{\rm sims}$ is the standard deviation of the distribution obtained from the isotropic simulations. The results have been summarized in Table \ref{Tab:D-353GHz}. For the set of four foreground maps with different S/N ratios that we have used in the above subsection, we find that the noisy foreground maps show large deviation from the isotropic simulations. We also find that for low S/N ratio, the $\mathcal{D}$ values of the noisy foreground maps increases.    

From these results, we can demonstrate two salient features of $\mathcal{D}$ statistic. First, $\mathcal{D}$ is quite robust in detecting the directional signal we are using, even if it is contaminated by isotropic white noise. Naturally, it's sensitivity towards the anisotropic signal reduces as the S/N ratio is reduced. However, it is significant enough for the level of noise present in the Planck polarization maps. Secondly, we note that the $\mathcal{D}$ value for the anisotropic map increases and approaches unity as the level of the isotropic noise is increased and the total map becomes more isotropic.
\vspace{0.1in}
\begin{table}
    \begin{centering}
    \begin{tabular}{|c|c|c|c|c|}
    \hline 
     & FG only & $S/N=2.0$ & $S/N=1.0$ & $S/N=0.5$\tabularnewline
    \hline 
    \hline 
    $\bar{\mathcal{D}}_{\rm sims}$ & $0.989 \pm 0.004$ & $0.993 \pm 0.002$ & $0.994 \pm 0.002$ & $0.994 \pm 0.002$\tabularnewline
    \hline
    \hline 
     $\mathcal{D}$ & 0.45 & 0.86 & 0.94 & 0.95\tabularnewline
    \hline 
    $\Delta \mathcal{D}$ & 135.2 & 62.7 & 32.6 & 21.9\tabularnewline
    \hline 
    \end{tabular}
    \par\end{centering}
    \caption{Table summarizing the values of $\mathcal{D}$ and $\Delta \mathcal{D}$ for different noise levels.}
    \label{Tab:D-353GHz}
\end{table}


\section{Statistical Isotropy of Planck data}
\label{sec:results}
We carry out our stated goal of testing SI using Planck data in this section. We compute $\alpha$ and $\mathcal{D}$ from Planck observations and simulations and compare the corresponding values. Agreement between the corresponding values computed from observations and simulations will imply that the observed maps are statistically isotropic.

\subsection{Statistical Isotropy of CMB \texorpdfstring{$E$}{E}-mode signal using CMTs}
\label{sec:cmbE}

\begin{figure}
    \centering
    \includegraphics[width=5.2in,height=4.2in]{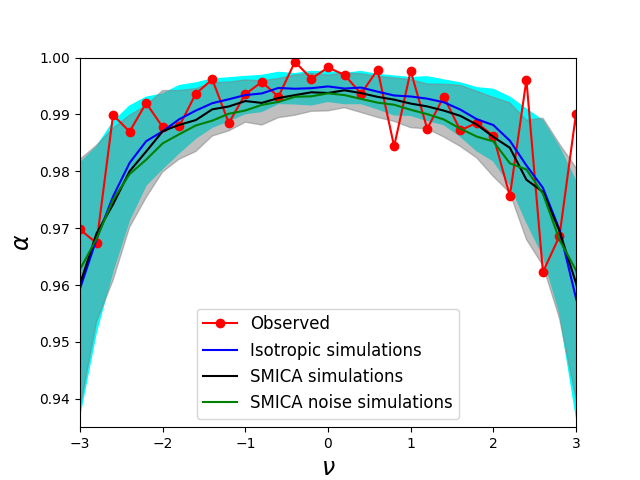}
    \caption{$\alpha$ values from the observed $E$ map (red line), mean (blue line) and 1-$\sigma$ limits (cyan shaded region) from 300 isotropic simulations with the same power spectrum, mean (black line) and 1-$\sigma$ limits (grey shaded region) from 300 SMICA simulations, and mean $\alpha$ from 300 SMICA noise simulations (green line).}
    \label{fig:alpha_iso}
\end{figure}

First, we check the consistency of $\alpha$ values computed from the observed $E$ map with those obtained from the 300 SMICA simulations, which include instrument noise effects. As mentioned in section ~\ref{sec:data}, the PLA provides the observed $E$ maps and 300 simulations at a resolution of $N_{\rm side}=2048$ and an effective beam of $5'$. We downgrade the pixel resolution of all the maps to a resolution of $N_{\rm side}=512$ and smooth them with a Gaussian beam of $20'$ FWHM, and then mask them with the common polarization mask before computing $\alpha$ from the maps. The results of our analysis of the SI of the $E$-mode maps are presented in figure ~\ref{fig:alpha_iso}. The $\alpha$ values computed from the observed $E$ map are represented by red dots. The black line and grey shaded regions denote respectively the mean and $1\sigma$ error bars from SMICA simulations. The green line denotes the mean $\alpha$ from SMICA noise simulations, the blue line and cyan shaded regions denote respectively the mean and $1\sigma$ error bars from 300 isotropic simulations. Conclusions drawn from comparison of $\alpha$ values for a single threshold value are unreliable due to statistical errors in the estimation of $\alpha$. Hence, we base our conclusions on the average behaviour of $\alpha$ in a range of thresholds. The 1$\sigma$ limits on $\alpha$ are lowest (high statistical significance) for thresholds close to 0, and they increase as we move away from 0 on either side. For our analysis, we choose the threshold range $\nu = -2$ to $+2$, and on averaging over this range, we find that only 40.1\% of the SMICA simulations have $\alpha$ values greater than the corresponding values estimated from the observed map. This translates to a statistical significance of less than $1\sigma$ and therefore, we infer that the observed map is consistent with the SMICA simulations. However, since the observed map is noise-dominated \cite{planck18_vii}, we can not conclude on the SI properties of the CMB $E$-mode signal based solely on this comparison.

To test the level of SI in the CMB $E$-mode signal, we compare the observed $E$ map with 300 isotropic simulations generated using HEALPix. We perform this analysis using maps with pixel resolution $N_{\rm side}=512$ and with a smoothing beam of $20'$ FWHM. The observed map has CMB convolved with $5'$ FWHM Gaussian beam and instrument noise and so the simulations must also have the same components.  We compute $\alpha$ from the 300 isotropic simulations and compare them with $\alpha$ from the observed $E$ map and those from the 300 noise simulations. From visual inspection of figure ~\ref{fig:alpha_iso}, we note that both the observed $E$ map and the 300 noise simulations are consistent ($<1\sigma$) with the isotropic simulations. Additionally, for almost all threshold values in the range, $-2\leq \nu \leq +2$, the $\alpha$ values from the observed map (CMB+noise) are greater than the corresponding values from the noise simulations. That is, the addition of the CMB $E$-mode signal to the noise maps, leads to an increase in the $\alpha$ values of the resultant map. This means that the CMB $E$-mode signal is more isotropic than the noise simulations. Given that noise simulations are consistent with the isotropic simulations at $<1\sigma$, we infer that the  CMB $E$-mode  signal is consistent with SI at $1\sigma$.

\vspace{0.1in}

\subsection{Probing SI of CMB \texorpdfstring{$E$}{E}-mode signal at different angular scales}
\label{sec:needlets}

To understand the SI properties of the observed $E$ map at different scales, we perform the same analysis as in section ~\ref{sec:cmbE}, but on maps generated by applying needlet filters to the observed and simulated $E$ maps. Needlets are a type of spherical wavelets which do not require any tangent plane approximation \cite{narcowich:2006,marinucci:2007} and are naturally embedded into the manifold structures of a sphere. They are axisymmetric functions and are expressed in terms of spherical harmonics. By construction, they are perfectly localized in harmonic space, and can be made highly localized in pixel space as well by using suitable choice of the filter involved in the construction. As done in \cite{basak:2012,basak:2013}, selecting a band of multipoles, and using cosine filters peaking at specified band centers and approaching 0 towards specified minimum and maximum multipoles, yields maps that carry information in the range of scales corresponding to the selected band. By analyzing these needlet filtered maps for a set of bands, we can study the SI properties of the input map at different scales. We use the band centers, $\ell_{\rm center}$ = 0, 100, 200, 300, 400 and 600. For each band, the filters $f_{\ell}$ are cosine functions of the multipole $\ell$ which peak at the respective $\ell_{\rm center}$  and fall to zero at the nearest band centers on either side of $\ell_{\rm center}$. These are given by,

\begin{equation}
 \centering
   f_{\ell} = \left\{ \begin{tabular}{cl}
        $\cos{\left(\frac{\pi}{2} \cdot \frac{\left(\ell - \ell_{\rm center}\right)}{\left(\ell_{a}-\ell_{\rm center}\right)} \right)}$ & \text{if } $\ell$ $\leq$ $\ell_{\rm center}$, \\
         & \\
        $\cos{\left(\frac{\pi}{2} \cdot \frac{\left(\ell - \ell_{\rm center}\right)}{\left(\ell_{b}-\ell_{\rm center}\right)} \right)}$ & \text{if } $\ell$ > $\ell_{\rm center}$,
   \end{tabular} \right.
\end{equation}
where $\ell_{\rm center}$ denotes the band center of the given band, and $\ell_a$ and $\ell_b$ denote the band centers of the previous and the following bands, respectively. For the bands centered at $\ell_{\rm center}$ = 0, 100, 200, 300, 400 and 600, the needlet maps are generated by multiplying the $a_{\ell m}$s of the input map with $f_{\ell}$, at $N_{\rm side}$ = 32, 64, 128, 128, 256 and 256 respectively.

\begin{figure}
    \centering
    \resizebox{3.0in}{2.0in}{\includegraphics{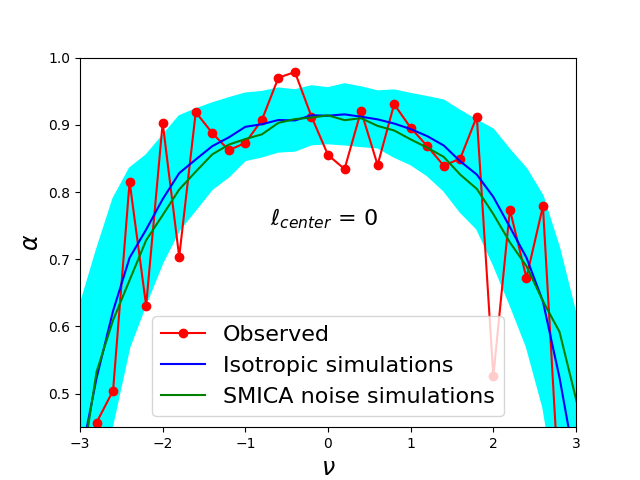}}
    \hskip 0in \resizebox{3.0in}{2.0in}{\includegraphics{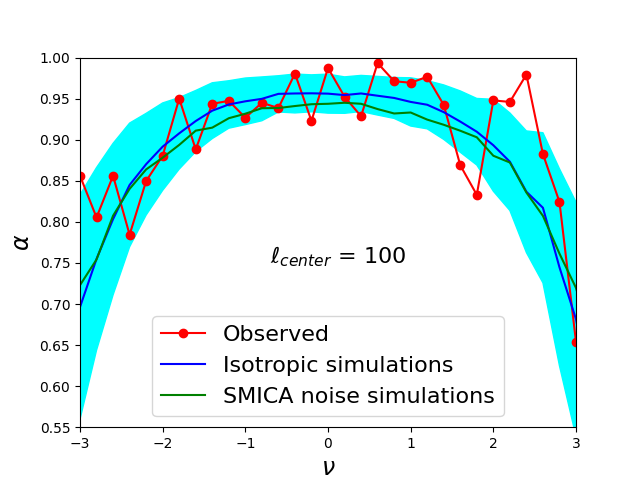}}
    \vskip 0in
    \resizebox{3.0in}{2.0in}{\includegraphics{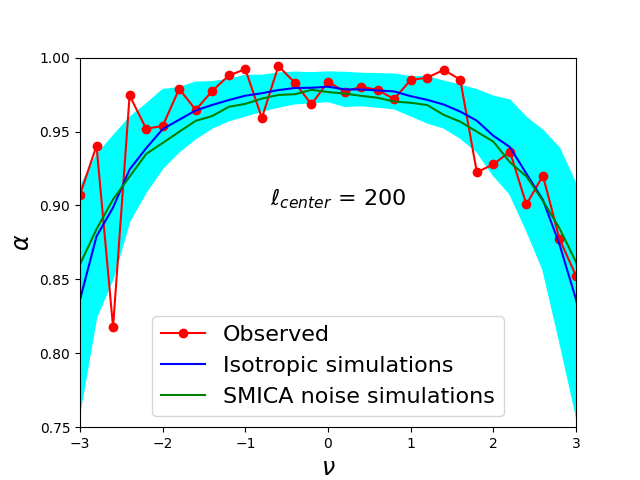}}
    \hskip 0in \resizebox{3.0in}{2.0in}{\includegraphics{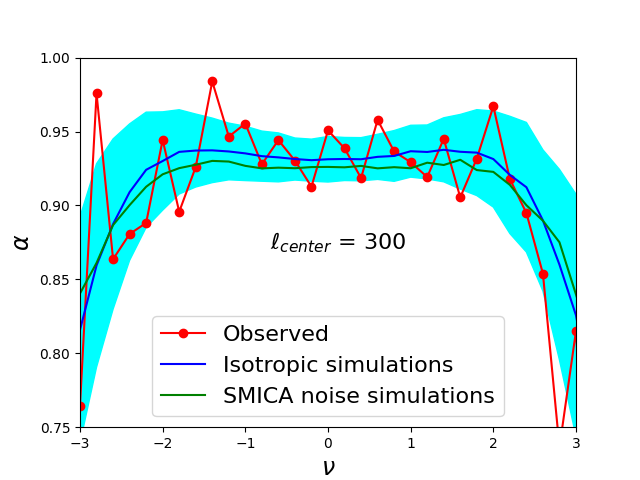}}
    \vskip 0in
    \resizebox{3.0in}{2.0in}{\includegraphics{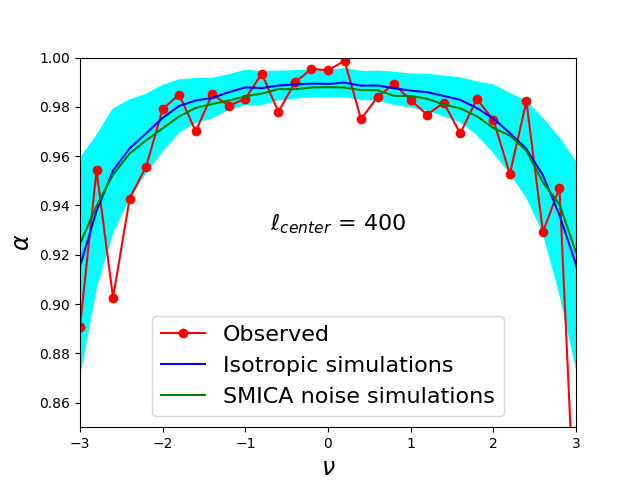}}
    \hskip 0in \resizebox{3.0in}{2.0in}{\includegraphics{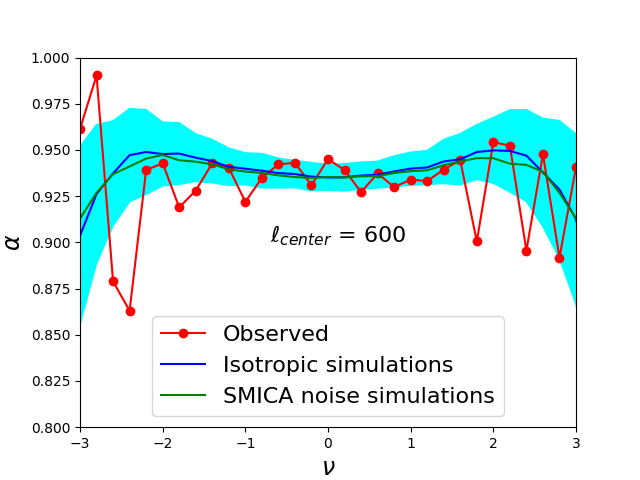}}
    \caption{Comparison of $\alpha$ from observed $E$ map (red lines), SMICA noise simulations (green lines), and isotropic simulations (blue lines) with $1\sigma$ limits (cyan shaded regions), for the needlet filtered maps with bands centered at $\ell = 0$, 100, 200, 300, 400 and 600.}
    \label{fig:needlets}
\end{figure}


The outcome of the needlet analysis is shown in figure ~\ref{fig:needlets}. The red lines represent the observed map, the green lines represent the mean of 300 noise simulations while the blue lines with cyan shaded $1\sigma$ regions represent the 300 isotropic simulations. To infer the SI property of each of the needlet filtered maps, we focus on the threshold range, $\nu = -2$ to $+2$. Across all the scales, both the observed map and the noise simulations are consistent with isotropic simulations within the $1\sigma$ limits. Additionally, with the exception of $\ell_{\rm center} = 600$ band, the $\alpha$ values for the observed map are greater than those for the noise simulations for most of the threshold values from $-2$ to $+2$. Hence, using a similar argument as presented in section ~\ref{sec:cmbE}, we can conclude that the CMB $E$-mode signal is statistically isotropic across all the scales. For the $\ell_{\rm center}$ = 600 band, it is not clear as to why the observed $E$ map $\alpha$ values are comparable to the noise simulations $\alpha$ values. Further investigation which is beyond the scope of this article, would be required to draw physical inferences from this result.

\subsection{\texorpdfstring{$\mathcal{D}$}{D} statistic analysis for SI of Planck polarization data}
\label{ssec:D-Stat Comp}
Next, we present the results  obtained by using $\mathcal{D}$ statistic from the Planck data. In order to draw conclusions about the SI of the signal in the data we have compared $\mathcal{D}$ statistic results from the observed $E$ map with four different sets of simulations. All the specifications relevant to these data and simulations have been described in section ~\ref{sec:data}. In order to mitigate the directional contributions from the foreground residuals present in the Galactic plane, we have applied the common polarization mask prior to applying $\mathcal{D}$ statistic on these data set and simulations.

\begin{figure}
\centering
    \resizebox{6.0in}{4.0in}{\includegraphics{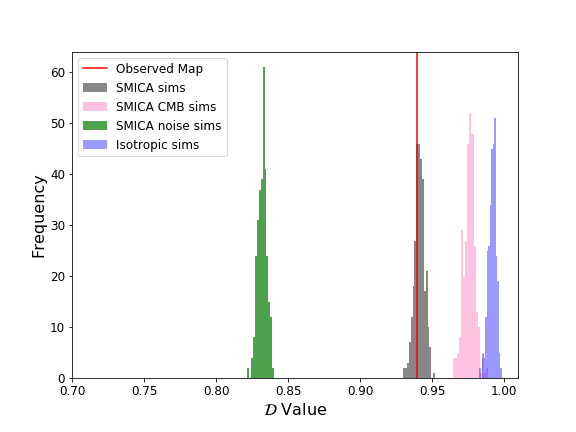}}
     \caption{The $\mathcal{D}$ values obtained from different sets of simulations and the observed $E$ map have been compared. Both simulation and data have been masked using the common polarization mask. The $P_p$ quantity have been set to 1 to obtain the pixel weights. Hence the $\mathcal{D}$ values calculated here are independent of any set of simulated maps.}
    \label{fig:D_comp}
\end{figure}

The results obtained from our analysis are displayed in figure ~\ref{fig:D_comp}. We find that the $\mathcal{D}$ value from the observed map is statistically consistent with the histogram obtained from the FFP10 SMICA simulations. This result concurs with earlier work \cite{D.Ghrear et al}, where the authors have implemented $\mathcal{D}$ statistic on Polarization angle maps. Since FFP10 SMICA simulations contain both the CMB $E$-mode signal and the noise, we use the isotropic simulations (described in \ref{sec:cmbE}) as our reference to probe for the SI of the actual signal. Our result clearly shows that the $\mathcal{D}$ value obtained from the observed map is not consistent with the distribution from the isotropic simulations. However, we have not taken into account the impact of frequency-dependent Rayleigh scattering, Doppler boosting and asymmetric beams in our isotropic simulations \cite{planck18_vii}. In order to take these effects into consideration, we have used the SMICA CMB simulations. Results obtained from these simulations show that, the SMICA CMB simulations have lower range of $\mathcal{D}$ values compared to the isotropic simulations. Nevertheless, the $\mathcal{D}$ value from the observed $E$ map still show large deviation from the mean of the SMICA CMB distribution. To investigate this issue further we use the SMICA noise simulations. The $\mathcal{D}$ values from these simulations indicate that they have significant level of anisotropy compared to the data as well as the SMICA CMB simulations. We already know that the noise $EE$ spectrum is comparable to the CMB $EE$ spectrum in the Planck data across all scales (described in \ref{sec:noise_level}). Hence, the presence of the anisotropic noise even if our signal is isotropic can make the resultant data anisotropic. We have also demonstrated in section ~\ref{sec:noisy} that the $\mathcal{D}$ statistic is sensitive enough to pick up on such anisotropy. 

From the above findings we deduce that the CMB signal present in the observed $E$ map is certainly more isotropic than the Planck noise. This conclusion is also in agreement with our findings using CMTs. However, contrary to the results from the CMT analysis, we find that the $\mathcal{D}$ from observed map is not consistent with SI. We already know that the $\mathcal{D}$ statistic is more sensitive to anisotropy than $\alpha$, at large scales. 
This have been further demonstrated in Appendix~\ref{App:HA} using hemispherically anisotropic map simulations. 
Hence, we conclude that the anisotropy in the data is at large scales. Since the Planck observations have well documented large-scale systematics, this is the most likely source of the anisotropy which is being detected by  $\mathcal{D}$ statistic.
\section{Conclusions}
\label{sec:conc}

The Contour Minkowski Tensor and $\mathcal{D}$ statistic provide independent tests of the SI of random fields in real space. The two methods are complimentary in the sense that the CMT is sensitive to small-scale information, while $\mathcal{D}$ statistic is more sensitive at large-scales. For a given random field, the ratio, $\alpha$, of the eigenvalues of the CMT, $\mathcal{W}_1$, quantifies the level of alignment of structures in the level sets of the field. How far $\alpha$ lies from unity, quantifies the level of alignment of the structures in the field, and hence is a measure of the deviation from SI in the field. Similarly, for $\mathcal{D}$ statistic the preferred directionality in the field is measured in terms of the alignment of the gradient vectors defined for the field. For an isotropic map the $\mathcal{D}$ value should be close to unity. Deviation from unity signifies the presence of anisotropy. For a pixelated map, $\alpha$ is never equal to one even for a statistically isotropic map. For this reason, we compare the $\alpha$ values computed numerically from observed maps with those computed from corresponding simulations which are statistically isotropic. In this work, we have applied the CMT and $\mathcal{D}$ statistic techniques to the observed $E$ map. We computed $\alpha$ from the observed $E$ map and from the FFP10 simulations. We find that both $\alpha$ and $\mathcal{D}$ statistic for the observed $E$ map are consistent with those from the FFP10 simulations. However, the observed map is very noisy and has a signal-to-noise ratio of $\approx 1$, and so it is not easy to quantify the SI of the CMB $E$-mode signal. To understand the effect of the addition of an isotropic component to a given map, on the two statistics used here, we compare $\alpha$ and $\mathcal{D}$ statistic values computed from anisotropic foreground simulations with varying levels of added isotropic white noise, with those computed from isotropic simulations. We find that the addition of an isotropic field to an anisotropic signal, increases the level of isotropy of the resultant map. We make use of this result to infer the SI properties of the CMB $E$-mode signal.

We compare the $\alpha$ and $\mathcal{D}$ statistic values computed from the observed $E$ map with those computed from isotropic simulations having the same power spectrum, and those from noise simulations. We find that $\alpha$ computed from all three sets of maps are consistent with each other at $< 1\sigma$. The $\alpha$ values for the observed $E$ map  containing the CMB signal and noise, are higher than the noise simulations. This tells us that the level of statistical anisotropy present in the CMB signal, is less significant than that present in the noise simulations. Since the noise simulations are consistent with the isotropic simulations at $1\sigma$, it implies that the CMB signal is consistent with the assumption of SI, at $< 1\sigma$. However, from $\mathcal{D}$ statistic we find that the observed map is not consistent with the isotropic simulations. Since, we have not considered the effects of asymmetric beams in generating the isotropic simulations, we use the more realistic SMICA CMB simulations instead. Even after comparing with these simulations, $\mathcal{D}$ statistic indicates the presence of anisotropy in the data. As $\mathcal{D}$ statistic is sensitive at large-scales, we conclude that the anisotropy is present at large-scales. But this does not necessarily mean that CMB is statistically anisotropic at large-scales. The the $E$ mode data is noisy across all scales and the Planck polarization maps have well documented large-scale systematic effects. This is the most likely source of the anisotropy which is being detected by $\mathcal{D}$ statistic.

\acknowledgments
Joby would like to thank NISER Bhubaneswar for the institute postdoctoral fellowship. All the Minkowski Tensor computations in this paper were run on the Aquila cluster at NISER supported by Department of Atomic Energy of the Govt. of India. TG acknowledges support from the Science and Engineering Research Board of the Department of Science and Technology, Govt. of India, grant number \texttt{SERB/ECR/2018/000826}. P. C. acknowledges support from the Science and Engineering Research Board of the Department of Science and Technology, Govt. of India, grant number \texttt{MTR/2018/000896}. AS would like to thank Majd Ghrear for some helpful discussions on $\mathcal{D}$ statistic. Some of the results in this paper have been derived using the HEALPix \cite{Gorski:2005,HPX} package. We acknowledge the use of the CMB polarization data provided by the Planck mission which is funded by the ESA member states, NASA, and Canada.

\appendix
\section{Hemispherical Anisotropy}
\label{App:HA}
In order to understand the differences/similarities b/w the two SI tests, we test their sensitivity towards the hemispherical anisotropy (HA). We use a very simple model for this,
\begin{equation}
    T(\hat{\boldsymbol{n}})=T(\hat{\boldsymbol{n}})[1+A(\hat{\boldsymbol{n}}.\hat{\boldsymbol{k}})].
    \label{eq:HA model}
\end{equation}
We have simulated temperature maps from Planck best-fit $TT$ power spectrum at $N_{\rm side}= 512$. We have introduced HA in the maps using the toy model in eq.~\eqref{eq:HA model}. The maps have been smoothed with $20'$ FWHM Gaussian beam. Both $\mathcal{D}$ statistic and MT analysis have been applied to these maps. To measure the sensitivity of $\mathcal{D}$ statistic towards HA we define a quantity $\delta \mathcal{D}$ for each map as follows,
\begin{equation}
    \delta \mathcal{D} = \frac{\mathcal{D}_I-\mathcal{D}_A}{\mathcal{D}_I}.
    \label{eq:delta D}
\end{equation}
Here, $\mathcal{D}_I$ is the $\mathcal{D}$ value from an isotropic map while $\mathcal{D}_A$ is derived from the anisotropic map. We expect to see a larger value of $\delta \mathcal{D}$ as we increase the level of anisotropy in the maps. 
In order to compare with $\mathcal{D}$ statistic which is defined over the complete sky, we integrate $\alpha$ over all threshold values. We have used the ``Simpson's Rule" from the {\em scipy} library to integrate over the threshold range of $[-2,2]$. We have normalized the integral to obtain a single parameter for the full sky. Using this quantity we define $\delta \alpha$ for an anisotropic map which is similar to $\delta \mathcal{D}$.

\begin{figure}
    \centering
    \resizebox{4.5in}{3in}{\includegraphics{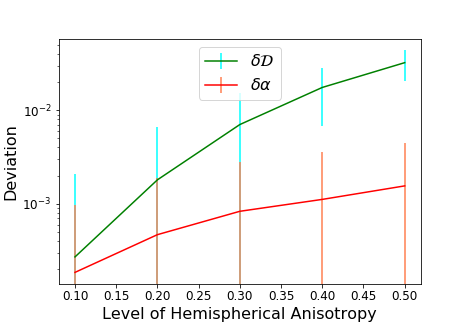}}
    \caption{The variation of $\delta \mathcal{D}$ and $\delta \alpha$ with HA. The mean value from 100 map simulations with $N_{\rm side}=512$ and FWHM=$20'$ have been plotted. The error bars are given by standard deviation from the 100 maps.}
    \label{fig:HA_Compare_ns512_f20}
\end{figure}

Figure ~\ref{fig:HA_Compare_ns512_f20} depicts our results. We have plotted the mean of $\delta \mathcal{D}$ and $\delta \alpha$ from 100 simulated maps. The error bars have been obtained from the standard distribution of $\delta \alpha$ and $\delta \mathcal{D}$. We find that $\mathcal{D}$ statistic has sensitivity towards HA while $\alpha$ is not sensitive towards it. As HA is a large-scale phenomenon this result is expected.


\end{document}